\documentclass[epj, referee]{svjour}
\usepackage{graphicx,amsmath}
\usepackage{bm}
\begin{document}
\title{Multiscale approach to radiation damage induced by ion
  beams: complex DNA damage and effects of thermal spikes}
\author{E. Surdutovich\inst{1,
    2}\thanks{\emph{E-mail:surdutov@oakland.edu; Tel:+1-248-370-3409}}
  \and A. V. Yakubovich\inst{1,3} \and
  A. V. Solov'yov\inst{1,
    3}\thanks{\emph{E-mail:solovyov@fias.uni-frankfurt.de}}
} 

\institute{Frankfurt Institute for Advanced Studies,
  Ruth-Moufang-Str. 1, 60438 Frankfurt am Main, Germany \and
  Department of Physics, Oakland University, Rochester, Michigan
  48309, USA \and On leave from
    A.F. Ioffe Physical-Technical Institute, 194021 St. Petersburg,
    Russia}

\date{\today}
%
\abstract{We present the latest advances of the multiscale approach to
  radiation damage caused by
  irradiation of a tissue with energetic ions and report the most
  recent advances in the calculations of complex DNA damage and the
  effects of thermal spikes on biomolecules.  The
  multiscale approach aims to
  quantify the most important physical, chemical, and biological
  phenomena taking place during and following irradiation with ions
  and provide a better means for clinically-necessary calculations
  with adequate accuracy.
We suggest a way of quantifying the complex clustered damage, one of
the most important features of the radiation damage caused by
ions. This method can be used for the calculation of irreparable DNA damage.
We include  thermal spikes, predicted to occur in tissue for a short time after
ion's passage in the vicinity of the ions' tracks in our previous
work, into modeling of the thermal
 environment for molecular
 dynamics analysis of ubiquitin and discuss the first results of these
simulations.}
\maketitle
\section{Introduction}
\label{intro}
{}
The success of heavy-ion-beam therapies, employed in Germany and
Japan, stems from several advantages of these therapies over
the common photon
therapies~\cite{Kraft05,Kraft07,Hiroshiko}.
These advantages can be described in the following way.
First, the Bragg peak in the linear energy transfer (LET) dependence on the
penetration depth gives an opportunity to better localize the dose
distribution on the targeted area. Provided that the targeted radiation damage
requires this dose, the overall delivered dose in ion-beam therapy is
smaller. This makes the ratio of the the doses delivered by photons to
that of ions (the (overall) relative biological
effectiveness (RBE)) larger than one (if there is no significant
overkill effect in the Bragg peak region). This advantage is
substantiated by the possibility of achieving relatively sharp edges in
the dose distribution, which spares vital organs, not touched by the
tumor, from irradiation, thus reducing side effects.
Second, the concentration of radiation damage caused by high-LET ion
irradiation is significantly larger than that of photon
irradiation. This changes the radiation damage not only quantitatively
(by increasing the dose localization) but qualitatively as well, i.e.,
the pathways of radiation damage change so that the direct effects
dominate the indirect ones. This solves the problem of cell
resistivity to irradiation and increases the local RBE, even for
hypoxic tumors.

Despite the successes of ion therapies there are many unanswered
questions. The
scenario of events from the incidence of an ion onto tissue to the
cell death is vague. Some important processes are
not understood even on a qualitative level. Given that the radiation damage
to DNA is mostly responsible for
cell death~\cite{Kiefer,Kanaar,Frankenberg,Goodhead93}, the pathways
of this damage are not sufficiently quantified. The roles of different
factors are still being evaluated. The approaches to
calculating the local RBE, like the Local Effect
Model~\cite{Kraemer,Elsaesser,LEMcluster}, which is based on local
dose with an ad-hoc 
accounting for complex damage, may be sufficient for now; but,
the future of ion-beam therapies requires a more sound phenomenological
(if not an {\em ab
initio}) calculation of the RBE.
The main
obstacle to understanding
radiation damage to DNA on the microscopic level is that the scenario
includes events on
many spatial, temporal, and energetic scales; e.g., time scales for
relevant processes vary from 10$^{-22}$~s to minutes, hours, or even
longer times. Indeed, 10$^{-22}$~s is the characteristic time of
nuclear reactions, which take place when an incident ion collides with
nuclei of the medium; 10$^{-17}$~s is that of ionization and
excitation of molecules of the medium, which are the leading processes
of energy loss by the projectile, 10$^{-12}$~s is that of  the transport of
secondary electrons  formed as a result of the
above ionization, 10$^{-5}$~s is that of DNA damage, and longer times
correspond to DNA repair by different mechanisms. These scales are
presented in Table~\ref{circle}.

The claim of our multiscale approach to the physics of
ion-beam cancer therapy is that the phenomenon-based calculation of
the RBE is possible if we evaluate the most important physical, chemical, and
biological effects that happen in the process of irradiation and
(mainly biological) processes following irradiation on longer time
scales. Instead of reconstructing the sequence of events using
scale-dependent
Monte
Carlo (MC) simulations, we consider phenomena on all scales and
combine them in a complete
picture~\cite{nimb,epjd,RADAM2,pre,epn,preheat,Scif}.

The understanding of the scenario of DNA damage and repair is an
interdisciplinary science problem, and its whole scope is
shown in Table~\ref{circle}.
\begin{table*}
\caption{Disciplines and scales of ion-beam cancer therapy}
{\begin{tabular}{llll}
    \hline\noalign{\smallskip} Phenomenon &Discipline & Space scale &
    Time scale\\
\noalign{\smallskip}\hline\noalign{\smallskip}
Beam generation & High energy physics &	m--km & \\
Beam transport & Radiation physics & 1--100 cm &$10^{-7}$s\\
Nuclear collisions and fragmentation &	Nuclear physics	& fm &
$10^{-22}$s\\
Primary ionization, transport of secondaries &	Atomic/molecular
physics & 0.1--10 nm &	$10^{-17}-10^{-12}$ s\\
Branching of secondaries, radicals,& Chemistry & 1--10 nm &
$10^{-12}-10^{-5}$s \\
 excited species, chemical equilibrium &{} &{} &{} \\
Local heating, heat transfer, stress & Thermo/hydro-dynamics & 1--10
nm & $10^{-14}-10^{-9}$s\\
Dissociative electron attachment to molecules &
Quantum chemistry & \AA & $10^{-15}$s\\
 and other reactions &{} &{} &{} \\
Initial damage effects & Biochemistry &	0.1--10 nm & $10^{-5}$s\\
Repairing mechanisms &	Molecular biology & 1--100 nm &	s--min \\
Cellular network and interaction & Cell biology & $\mu$m & min \\
(Tumor) Cell death & Medicine &	mm &	min--years \\
\noalign{\smallskip}\hline
\end{tabular}}
\label{circle}
\end{table*}
From this table, one can see that this problem joins different areas of
physics, chemistry, and biology.
This scope is too vast for taking on all scales
  simultaneously, and in the beginning we
limited our considerations to physical and some chemical
phenomena. At this moment, our multiscale approach consists of
analyses of ion propagation in a medium, production and transport of
secondary electrons, and different pathways of DNA damage and their
quantification.

We continue this paper with a quick overview of ion transport and
production of secondary electrons in Section~\ref{sec:ions}. Then we
address several issues of DNA
damage in Section~\ref{pathways}. Two subsections in that section, on
cluster damage and on the effect of thermal spikes on ubiquitin, are
new steps in our development of the multiscale approach.

\section{Ion stopping and production of secondary electrons}
\label{sec:ions}
Energetic ions incident on tissue lose energy primarily through
ionization of molecules of the medium. The main characteristic, which
describes the phenomena on this scale (by far the longest spatial) is
the singly differential (in the energy of released electrons) cross
section of ionization (SDCS)~\cite{epjd}. This cross section is
proportional to the
energy distribution of secondary electrons, which is, in its turn,
important for the calculation of DNA damage. For the ion's transport,
the first moment of the SDCS along with the excitation cross section
contributes to the total energy loss or stopping cross section. For many
purposes, the LET,
inversely proportional to the stopping cross section, is a more
convenient characteristic. Most important, the LET dependence on the
depth in tissue gives the longitudinal dose distribution, which is
directly related to the radiation damage. The Bragg peak in the LET
dependence on depth, a sharp maximum close to the end of the ion's
track, is a major contribution to the advantages of ion-beam
therapies.

The subsequent DNA damage is done either by the secondary
electrons, produced at
this stage~\cite{SevDQD,SevReview,SevGreen,Sanche05}, or by the holes
(also produced as a result of the ionization of
the medium); these comprise the direct and quasi-direct
effects~\cite{SevDQD,SevHole,SevReview,SevGreen}.  The
indirect effect is caused by hydroxyl radicals resulted from
ionization and excitation of the medium. The direct effects are more
important when tissue is irradiated with heavy
ions; their relative independence from the chemical properties of the medium
(such as the presence of oxygen) is particularly attractive for
treatment of
hypoxic tumors. The interaction of secondary electrons with DNA is
defined by their energy and number density at the location of the DNA
molecule. The transport properties, such as the mean free path
of electrons, are also energy-dependent. Therefore, the energy spectrum
of the secondary electrons is very much desired. It turns out, however,
that at low energies, this distribution, in a medium such as liquid
water, is neither
easy to calculate, nor measure experimentally. We have gradually
improved our approach to calculating these spectra in
Refs.~\cite{nimb,epjd,Scif}. In Ref.~\cite{Scif} we have
improved the parametric approach at low energies of projectiles and
analyzed different options at high energies.

Note, that the dose concentration does not yet mean the reduction of
the dose. Thus far, it is assumed that the damage is proportional to the
dose, i.e., in order to eradicate a desired percentage of cells in a
given region the same dose is required there, whatever  the
projectile is. However, due to dose localization in ion therapies, the
dose in the surrounding regions is smaller than that in photon
therapy; therefore the overall RBE increases.

\section{Pathways of DNA damage}
\label{pathways}


The localization of dose associated with the Bragg
peak results in a high number density of secondary electrons. This
results in complex DNA damage~\cite{Ward1,Ward2}. The complex
damage reduces the capabilities of proteins to repair the
damage. Thus, the damage induced in high-LET irradiation is
more lethal to the cells. This adds a new quality to the dose and
poses a question of what is more important, the energy or
complexity of damage. This quality enhances the RBE locally and makes
up still another 
advantage of ion-beam therapies. This advantage is more related to chemistry
and biology even though it is a consequence of physical
effects.

\subsection{Estimation of DNA damage done by secondary electrons}

In order to quantify the second feature of high-LET irradiation, one
needs to investigate the pathways and mechanisms of DNA damage and
repair. It is widely accepted that double-strand breaking (DSB) is
the most pernicious type of DNA damage. This damage is defined as
the breaking of both strands of DNA on the length of a single DNA
convolution. It is also known  that an isolated DSB can be
repaired; however if additional lesions occur nearby, i.e., if
damage is clustered sufficiently, it may be lethal for the cell. A
single-strand break (SSB) can easily be repaired; but, if
several of them happen close enough together (not causing a DSB) then the
so-called multiple strand break may also result in irreparable damage.

Using the results of Refs.~\cite{Dingfelder,Nikjoo06,Tung,Sanche05}, we
considered the diffusion of low-energy secondary electrons from the
place of their origin to the DNA~\cite{pre}. This work showed
that the probability of a particular DNA convolution to encounter a
secondary electron, is strongly dependent on the distance from the
convolution to the ion's path, and only slightly depends on the
orientation of the
convolution with respect to the ion's path. This result has let us make
an estimate of a number of DSBs caused by an ion's passage through a cell
nucleus. An estimate was made for carbon ions passing through glial
cells, which comprise 90\% of the human brain. It predicted
3.5~DSBs per
$\mu$m of the ion's track in the vicinity of the Bragg
peak~\cite{pre}. This number
can be very reasonably compared with the number, 2.6 DSBs per
$\mu$m, reported by experimental studies of
Taucher-Scholz et al. at GSI~\cite{Jacob}. However, these are only the
first crude estimates. They only consider one pathway of DNA
damage. For the probability of the DSB yield in an electron-DNA collision, we
relied on the experimental data of Sanche~\cite{Sanche05}. These
famous experiments, however, may not represent the whole picture,
because the DNA used in these experiments had not been hydrated
as it is {\em in vivo} and its properties, such as the probability of a
DSB
after being hit by an electron, may be different. More research is
required in order to better quantify this pathway. Other pathways
include damage done by holes formed in the process of ionization,
possible damage due to the temperature increase in the vicinity of the
beam, the damage done by radicals, and all possible combinations of
the above.

\subsection{Calculation of clustered damage}

As we have already pointed out, the damage complexity is one of the
consequences of
the so-called
high-LET irradiation because in regions where
the LET is high, many agents of damage are
produced. This increases the probability of several agents, such as
secondary electrons, holes, and hydroxyl radicals, to make lesions in
the adjacent regions of DNA. Such lesions, SSBs, DSBs, and base
damage, combined together on a distance of less than 100 bp make up
the complex damage sites. While biologists study the pathways of
repair of such sites~\cite{Lynn1,Lynn2}, biological physicists
investigate the
cause of such damage~\cite{Rob1,Rob2}. In this work, we want to quantify the
complex damage without MC simulations. Such a quantification
is an important part of our multiscale approach to radiation
damage.

Imagine an ion,
near its Bragg peak, passing through a cell nucleus. It ionizes the
tissue, and secondary electrons, holes, and formed radicals cause DNA
damage in the vicinity of the track. The damage may be clustered;
i.e., along the track there are some SSBs, DSBs, oxidative base
damages, abasic sites, and combinations thereof.  Then, within minutes,
the repair mechanisms become active: H2AX histone becomes phosphorylated
($\gamma$-H2AX) and e.g., attracts such repair agents as proteins 53BP1, NBS1
and MDC1. These $\gamma$-H2AX-centered aggregations, called ``foci'', are
observed and remain until the DNA is repaired~\cite{Jacob}.

In general, it may
not be easy to classify different cluster damage, because there are
too many different possibilities~\cite{Rob1,Rob2}. We suggest classifying
clusters by the number of independent agents, which cause the
damage. For simplicity we will assume, for now, that only secondary
electrons cause damage to DNA.  For example, if a single secondary
electron brings about a DSB and nothing else is damaged within certain
region, we will refer to such a site as a singly-damaged site
(DS1). If two electrons damage a DNA molecule within a certain
distance, we will refer to this as a doubly-damaged site (DS2). In
this case, the condition, which defines the maximum distance between
the lesions is that only one repair focus is formed to repair the
site. Then we can similarly introduce DS3, DS4, etc.

Thus, in order
to approach the clustered damage, we should only consider the lesions
due to a single electron, such as SSBs, DSBs, base damage, etc., with
certain energy-dependent probabilities. Then, since secondary
electrons are mutually independent, we can calculate the probability
of clustered damage, such as DS1, DS2, etc., using Poisson
statistics.

Let us consider one example of such a calculation for a
superficial case where each electron causes a DSB with a certain
probability and these DSBs may be clustered. This example is based on
the probability of DSBs caused by secondary electrons~\cite{pre}.
Let us
suppose that each DSB can be surrounded by some volume, such that, if
any other DSB occurs within this volume, it will be counted as a
member of the same cluster (e.g., if there are no other DSBs within
this volume then this will be a cluster containing just one DSB, i.e.,
DS1).
Let us denote this volume as $V_C$ . Let the probability of a single DSB
occurring inside this volume be $p(1)$. The probability of two DSBs
occurring within this volume is $p(2)$, that of three DSBs is $p(3)$,
etc.   Let us
now calculate these probabilities, taking the volume of the cluster $V_C$
to be the volume of one nucleosome bead consisting of eight histones
wrapped by a stretch of DNA consisting of 146 bp. This seems to be a
reasonable unit related to DNA geometry, and the number of bp involved
is of the order of the upper limit of the modeled cluster
damage~\cite{Rob1,Rob2}.

  According to the method of calculating the
number of DSBs per convolution volume~\cite{pre}, overviewed above,
 this number (for glial cells with a convolution right next to the
ion's track) is $N_{conv}\approx 3\times 10^{-3}$).  As in that work, we assume
that the DNA is uniformly distributed inside cell nuclei. This
assumption is reasonable for cells in their interphase, where they
spend most of their life. We can translate this number into an average
number of DSBs per cluster volume $N_C=N_{conv} n_{conv} V_C$, where
$n_{conv} = 4\times 10^6 \mu$m$^{-3}$ is the number density of
convolutions inside
the cell nucleus (for glial cells). Then $N_C=1.2\times10^{-2}$ (for
$V_C$=10-6 $\mu$m$^3$) and the probability of exactly $k$ DSBs to occur
in this volume, $p(k)$, is given by the Poisson distribution:
\begin{equation}
p(k)=\exp{(-N_C)}\frac{N_C^k}{k!}~.
\end{equation}
With
the above $N_C$, $p(1)= 1.2\times 10^{-2}$, $p(2)= 7.1\times 10^{-4}$, and so
on. Neither $p(1)$ nor $p(2)$ depends on time. If this comparison works,
it proves that the above picture of damage is correct and the average
number of DSBs per cluster can be inferred (in this idealized case)
from the ratio of these probabilities:
\begin{equation}
N_C=2\frac{p(2)}{p(1)}~.
\label{ncratio}
\end{equation}
Notice that the
volume of the cluster $V_C$ is many times smaller than the volume of a
focus, $\approx 1 \mu$m$^3$~\cite{Jacob}. The latter corresponds to the volume
occupied by all proteins engaged in the repair process. Our assumption
that the cluster volume is that of a nucleosome bead is not important
for the suggested analysis and the size of a cluster can be changed if
necessary. For example, let us take the cluster volume to be 100 times
larger than the volume of a nucleosome, then $N_C=1.2$, and corresponding
$p(1)$ and $p(2)$ are $7.8\times10^{-2}$ and $9.1\times 10^{-2}$,
respectively. The ratio given by eq.~(\ref{ncratio}) still holds and does not
alter the logic of the suggested method. However, if $V_C$ turns
out to be too large (~1$\mu$m$^3$) then the probability of having just one
or two lesions will be unreasonably small. In this case, the whole
concept of the clustered damage will have to be reconsidered, since
the number of DSBs per cluster will be in triple digits. In general,
it is the repair mechanism that defines the $V_C$, and thus, it may be
smaller than the volume of the nucleosome as well.

The reality is less
straightforward than the considered example for at least two reasons:
each electron may cause different types of lesions with different
probabilities, and there is more than one pathway of DNA damage. Both
of these complications can be overcome in the following way. In order
to describe different types of lesions due to different agents, we can
introduce the probabilities $p_{11}, p_{12}, p_{13},... p_{21},
p_{22}, p_{23},... p_{31},
p_{32}, p_{33},$ etc., of certain types of lesion caused by a certain agent,
respectively. Each of these probabilities is similar to $N_C$ introduced
above. Then
\begin{eqnarray}
\nonumber
p(1)&=&\sum_{i, j}p_{ij}\exp{(-p_{ij})},~\\ 
\nonumber
p(2)&=&\frac{1}{2}\left[\sum_{i, j}p_{ij}^2\exp{(-p_{ij})}+\right.\\ 
\label{poneptwo}
&+&\left.\sum_{i, j\neq l}p_{ij}\exp{(-p_{ij})}p_{il}\exp{(-p_{il})}\right], 
\end{eqnarray}

\noindent and so forth. These are the probabilities of clustered
damage DS1, DS2, etc.  The above idea of classification of damage is
only attractive if the there are not too many terms in the
decomposition~(\ref{poneptwo})
to be considered. In order to make a judgment about this, we need
to analyze the difference between different lesions biologically. For
example, let a DSB be inflicted at a certain position in a DNA
molecule.  SSBs in the same molecule may make up a clustered damage
site if
they are located close enough to the DSB. Should we distinguish
between the situations when SSB is happening 10, 12, or 20 bp away
from the DSB? If biologists answer this and similar questions, we
can justify (or refute) the above approach.  In order to test the
suggested hypothesis, we suggest further study of the repair of
particular damage sites. We expect that the number of
effective configurations will be limited and the types of clusters
will be subdivided into two categories to justify the observed
biphasic repair dependence; those (not necessarily the simplest),
which are fixed at a fast rate (within two hours) and those (not
necessarily the most complex), which are fixed at a slower rate
($\approx 24$h).

\subsection{Effects of thermal and pressure spikes}

Now we return to the discussion of pathways of DNA damage.
In our works \cite{nimb,epjd}, we made
estimates for the temperature
increase in the vicinity of ion tracks. The temperature increase is
caused by secondary electrons that get most of the energy lost by the
stopping ions. Then, this energy is transferred to the medium as
electrons become thermalized and bound.  The temperature increase strongly
depends on the volume within which the energy is deposited. This
volume has been estimated using data on the penetration depth of
secondary electrons whose average energy is about 45~eV. The maximum
average temperature increase was estimated to be about
100$\,^{\circ}$C, which is
enough to denature DNA, but
these estimates have been done for a uniform system in thermal
equilibrium and the energy transport by electrons before a transfer to
the lattice has not been taken into account.

In Ref.~\cite{preheat},
we considered the heat transfer during the earliest stage after the ion's
passage. The characteristic times of this stage are from
$10^{-15}-10^{-9}$ s. These times are longer than the characteristic
time of primary ionization ($10^{-17}$ s) and shorter than typical times
of conformational changes in DNA such as unwinding, which are measured
in $\mu$s or even longer times. Nonetheless, the events that happen
on this intermediate time scale make the initial conditions for the next
scales and may be important for the future dynamics of the medium.

\begin{figure}
  \includegraphics[width=3.0in]{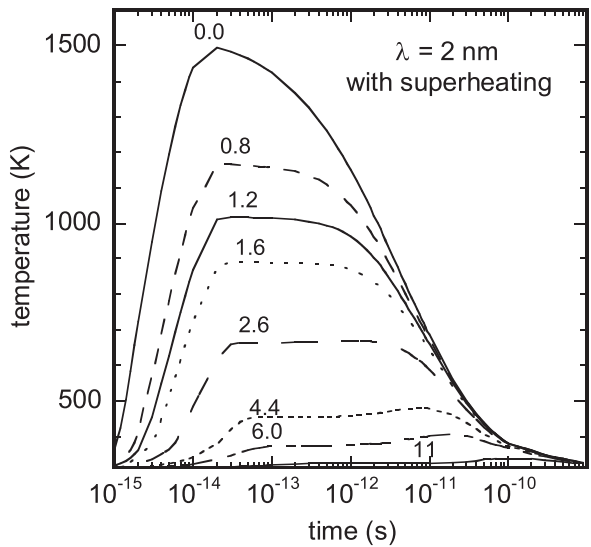}
\caption{The
  temperature on the molecular subsystem versus time for
one value of
the electron-phonon mean free path ($\lambda= 2$~nm),
 assuming a superheating scenario.
The calculations have been
performed for 0.5-MeV/u C ions (LET~$=0.91$~keV/nm~\cite{35}). The
calculations are performed for different radii
relative to the ion axis. These radii are given (in nm) near each
curve.
}
\label{fig2}
\end{figure}

We made calculations using the inelastic Thermal Spike model (i-TS),
which  was developed to explain
track formation in solids irradiated with heavy ions
\cite{Toul05,32}.
This
model studies the energy deposition to the medium by swift heavy ions
via secondary electrons. In this model, the strength of the energy
transfer from electrons to lattice atoms, called the electron-phonon
coupling, is an intrinsic property of the irradiated material.
We applied the i-TS model to irradiated liquid water.
Fig.~\ref{fig2}~\cite{preheat}
presents the temperature
of water versus time for a scenario with superheating and $\lambda =
2$~nm for a C-ion beam of 6 MeV or 0.5 MeV/u (LET~$=0.91$~keV/nm),
which corresponds to the energy of carbon ions in the vicinity of the
Bragg peak.  These results indicate a sharp increase of temperature
for a short time. This increase is {\em much} larger than has been
previously estimated in stationary conditions. During the times
between $10^{-15}$ and $10^{-10}$~s after the ion's passage, the
temperature rises considerably at different
distances from the ion track~\cite{preheat}.

The described system reaches thermalization only by the time of
  about $10^{-12}s$; therefore the temperature, which we discuss
  above, is rather a distribution of energy per molecule calculated in
  K. Nevertheless, the energy transfer inevitably takes place near ion
  tracks and the i-TS model presents a plausible picture of this
  transient process. A rapid energy transfer between internal degrees
  of freedom occurs on the femtosecond scale. Then, the energy is
  transferred to the translational degrees of freedom. If the i-TS
  model does not correctly reproduce the dynamics of this transition,
  the temperature spikes may be smaller.

Still, if real, very large
temperature and pressure increases (within 10~nm of the ion
trajectories, predicted in Ref.~\cite{preheat}) may result in
considerable forces acting on
DNA. These may be large enough to cause mechanical damage, such as
strand breaks, and thus be a separate mechanism of DNA
damage during irradiation by ions.

\subsection{Influence of the thermal spike on proteins structure}
\begin{figure}
\includegraphics[width=3.5in]{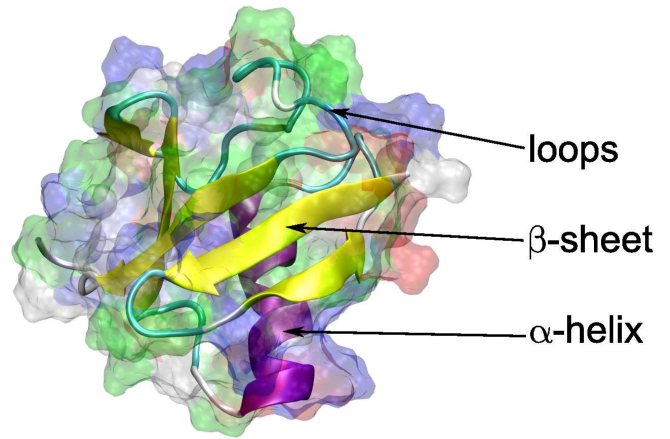}
\caption{The structure of protein ubiquitin (PDB ID 1UBQ) The Figure
  was rendered using VMD program~\cite{VMD}. The elements
  of the secondary structure are shown by arrows.}
\label{fg:ubiquitin}
\end{figure}

In order to investigate the stability of protein structures under
ultrafast heating events caused by the propagating heavy ions, we have
performed molecular dynamics simulations of proteins exposed to
heating events.  As a case study, we have chosen the small and well
known globular protein ubiquitin. The crystallographic structure of
this protein is shown in Fig.~\ref{fg:ubiquitin} and was obtained from
the Protein Data Bank~\cite{PDB} using ID code 1UBQ.  Ubiquitin
consists of 76 amino acids and its main function is associated with
the labeling of the proteins in the cell for proteasomal degradation.
The choice of ubiquitin as a test example is motivated by the fact
that this protein has helix and sheet elements of the secondary
structure, being one of the smallest and compact proteins.

The molecular dynamics simulations were performed using the following
procedure.  The protein with the crystallographic structure was
solvated in a water box with the edge size of 80~\AA. Water
was simulated using the TIP3 parameterization for water molecules.  The
solvent was ionized with ions of sodium and chloride with
concentrations of 100~mM.  The molecular dynamics simulations were
performed in the NVT ensemble using the NAMD software package~\cite{NAMD}
and the CHARMM27 forcefield~\cite{CHARMM}, and using a
timestep of 1 fs. The temperature control was
maintained using a Langevin thermostat with a damping coefficient of 5
ps$^{-1}$. In order to simulate the heating of the medium by
the energetic particle, the temperature of the thermostat was adjusted
after each 10 timesteps to the temperature profile shown in
Fig.~\ref{fig2}. It is important to mention that in the
presented calculations, we have not accounted for the spatial dependence
of the temperature peak and assumed that the whole system experiences
the temperature spike as if it is located on the trajectory of
the propagating particle.

We have performed four independent molecular dynamics simulations of
ubiquitin exposed to the same heating event. The simulations were
performed for the first 300~ps after the ion's passage. As it is seen
from Fig.~\ref{fig2}, the temperature
of the medium at 300~ps after propagation of the energetic particle is
lower than 350~K. Therefore, one can speculate that the prominent
changes in the secondary structure of the protein caused by the
temperature increase should occur during the first 300~ps after the
ion's passage, since, later on the temperature is rather low.

\begin{figure*}
\includegraphics[width=6.in,clip]{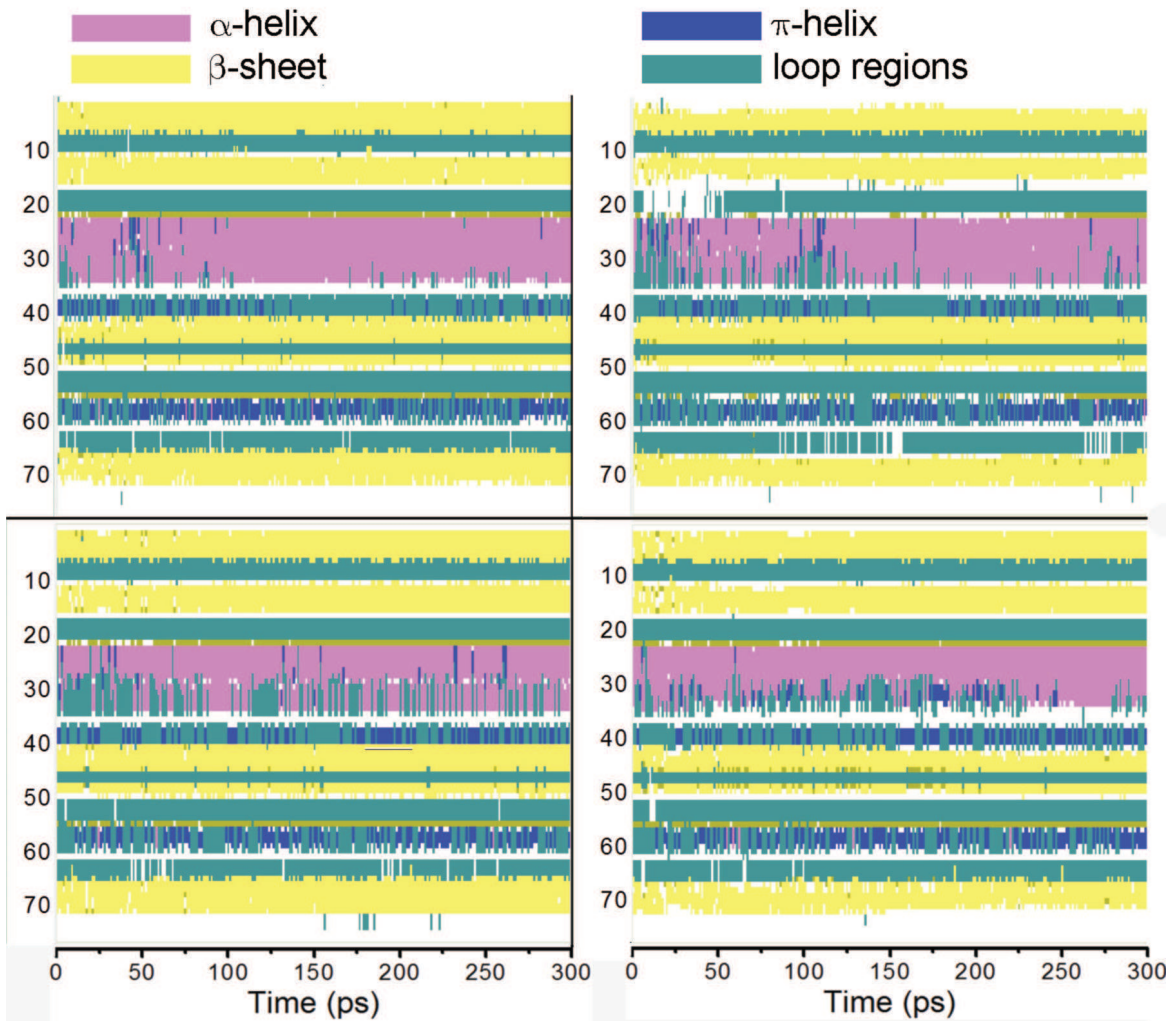}
\caption{The dependence of secondary structure of ubiquitin on time in
  a course of the heating event. The ordinate represents the index of
  amino acid in the protein. Each color corresponds to different types
  of secondary structure: purple represents $\alpha$-helices, yellow
  represents $\beta$-sheets, green represents loop regions and blue
  represents $\pi$-helices.  The white color represents the cases where it
  is not possible to assign any kind of secondary structure to the
  amino acids. The zero time corresponds to the moment of propagation
  of the energetic particle.}
\label{fg:protspike}
\end{figure*}

Since we are interested in the influence of the thermal spike on the
secondary structure of the protein, in Fig.~\ref{fg:protspike} we
present four dependencies of the secondary structure of ubiquitin on time
obtained from four independent calculations. On the vertical axis of each
plot in Fig.~\ref{fg:protspike} is the index of each amino acid in the
protein
while the horizontal axis is the time after the propagation of the
particle.  Each color represents different types of secondary
structure: purple represents $\alpha$-helices, yellow represents
$\beta$-sheets, green represents loop regions and blue represents
$\pi$-helices. The white color represents the cases where it is not
possible to assign any kind of secondary structure to the amino
acids. Thereby, from Fig.~\ref{fg:protspike} one can see which type of
secondary structure each amino acid has at a certain period of
time.

From Fig.~\ref{fg:protspike}, it is seen that the most prominent
disturbances of the secondary structure of the protein occur during
the first 100~ps after the propagation of the energetic
particle. However, during the time between 100 and 300 ps the
fluctuations of the secondary structure of the protein
decrease. Moreover, it is possible to state that at the
end of the simulation ubiquitin drifts back to its native
conformation since most of the elements of the secondary structure of
the protein are in the conformation corresponding to the native one.

Therefore, from the performed calculations it is difficult to
undoubtfully make conclusions about the influence of the thermal spike on the
secondary structure of the protein.  However, it is seen that the
secondary structure of the protein can be substantially distorted by
the thermal spike. But the proteins with fast folding kinetics such as
ubiquitin are able to rapidly refold to the native state. Therefore,
one can speculate that the most prominent effect of the damage on the
secondary structure of the protein should be observed in proteins with
a more complicated folding landscape. These kinds of proteins, which are not
able to refold themselves on their own, can be considered as the most
prone to damage. In the present work, we do not perform the
molecular dynamics simulations of proteins obeying slow and
complicated folding kinetics due to the large size of these systems
and we leave this question open for further considerations.

\subsection{Other effects of local heating}
Another issue which will have to be addressed is related to
  chemical changes in the DNA environment and their effects on DNA. It is
  expected that due to temperature spikes, the rates of chemical
  reactions increase by orders of magnitude. This concerns the
  dissociation of water as well as larger molecules. These
  processes produce more hydroxyl radicals and change the reactivity of
  the DNA environment. At the same time, DNA itself, which may be
partially
  denaturated, becomes more
  vulnerable to chemical damage.
Temperature spikes affect the probabilities of direct and quasi-direct
pathways of DNA damage by means of secondary electrons and holes since
the thresholds for some effects (such as vibronic
excitation)~\cite{SevET1,SevET2,SevHole} are
comparable to the energies transferred to the DNA via the heat conductance
mechanisms described in Ref.~\cite{preheat}. This means that
ionization of DNA with its concurrent heating may be the dominant
pathway leading to strand breaks.

\section{Conclusions and outlook}

In concluding this paper, we want to summarize our achievements in the
development of the multiscale approach to radiation damage and outline
some perspectives for future developments. First, on the stage of
propagation of ions we reproduced the position and shape of the Bragg
peak for protons and for carbon ions propagating in water. Several
important effects define this stage~\cite{epjd,Scif}: ionization and
excitation of the
medium, charge transfer, scattering, and nuclear fragmentation. The
latter has not yet been included in the multiscale approach and we
hope to do it in the future.
The energy spectra of secondary electrons produced during ionization
of the medium have been addressed in refs~\cite{epjd,Scif} and probably
can be further improved.
They serve as a starting point for the following stage of the transport of
 secondaries considered in ref.~\cite{pre} and heat transfer
 considered in ref.~\cite{preheat}. The transport stage can be further
 improved. The result should include the radial distribution of the
 clustered damage, described in this work. It is essential that the
 analysis of clusterization be related to investigations of DNA
 repair, which will one day also become a part of the multiscale
 approach. The heat transfer stage requires more efforts: the validity
 of application of the thermal spike model has to be verified. The
 direct effect on biomolecules should be further investigated using
 more realistic dynamical conditions. Finally, the other, indirect
 consequences  of thermal spikes have to be explored.
Thus, we have probably reached the ``end of the
beginning''; there is plenty of work yet to be done. We are hopeful
that a beautiful physical picture will finally describe the complicated
puzzle involved in the phenomena of ion-beam therapy.

\section*{Acknowledgments}
This work has contributed from discussions with Igor Mishustin, Walter
Greiner and
those at RADAM 2009.
E.S. is grateful to Lynn Harrison and J.~S. Payson. The
Deutsche Forschungsgemeinschaft support is very much appreciated.


\end{document}